\documentclass[12pt]{article}
\usepackage{graphicx}
\begin{document}
\renewcommand{\thefootnote}{\fnsymbol{footnote}}
\begin{titlepage}
\begin{flushright}
\end{flushright}

\vspace{10mm}
\begin{center}
{\Large\bf Noncommutative Extension of Minkowski Spacetime and Its Primary Application}
\vspace{16mm}

{\large Yan-Gang Miao\footnote{e-mail address: miaoyg@nankai.edu.cn}

\vspace{6mm}
{\normalsize \em Department of Physics, Nankai University, Tianjin 300071, \\
People's Republic of China}

\vspace{3mm}
{\normalsize \em Department of Physics, University of Kaiserslautern, P.O. Box 3049, \\
D-67653 Kaiserslautern, Germany}

\vspace{3mm}
{\normalsize \em Bethe Center for Theoretical Physics and Institute of Physics, University of Bonn, \\
Nussallee 12, D-53115 Bonn, Germany}}

\end{center}

\vspace{10mm}
\centerline{{\bf{Abstract}}}
\vspace{6mm}

We propose a noncommutative extension of the Minkowski spacetime
by introducing a well-defined proper time from the
$\kappa$-deformed Minkowski spacetime related to the standard basis.
The extended Minkowski spacetime is commutative, {\em i.e.} it is based on the standard Heisenberg commutation
relations, but some information of noncommutativity is encoded through the proper time to it. Within this framework,
by simply considering the Lorentz invariance we can construct field theory models
that comprise noncommutative effects naturally. In particular, we find a
kind of temporal fuzziness related to
noncommutativity in the noncommutative extension of the Minkowski spacetime. As a primary application, we investigate
three types of formulations of chiral bosons, deduce the lagrangian theories of noncommutative
chiral bosons and quantize them consistently in accordance with Dirac's method,
and further analyze the self-duality of the lagrangian theories in terms of the parent action approach.

\vskip 12pt
PACS Number(s): 02.40.Gh; 11.10.Nx; 11.10.Kk


\end{titlepage}
\newpage
\renewcommand{\thefootnote}{\arabic{footnote}}
\setcounter{footnote}{0}
\setcounter{page}{2}

\section{Introduction}
In history, Snyder~\cite{s1} published the first work on a
noncommutative spacetime although the idea might be traced back
earlier to W. Heisenberg, R.E. Peierls, W. Pauli, and J.R.
Oppenheimer. The quantized spacetime was introduced in order to remove the
divergence trouble caused by point interactions between matter and
fields. In modern quantum field theory, instead of a discrete
spacetime, the well-developed renormalization is utilized to
overcome this difficulty in the spacetime with the scale larger than
Planck's. However, the idea of noncommutative spacetimes has revived
due to the intimate relationship between the noncommutative field
theory (NCFT) and string theory~\cite{s2} and between the NCFT and
quantum Hall effect~\cite{s3}, respectively. In the former
relationship, some low-energy effective theory of open strings with a
nontrivial background can be described by the NCFT and thus some
relative features of the string theory may be clarified through the NCFT
within a framework of the quantum field theory, and in the latter the
quantum Hall effect can be deduced from the abelian noncommutative
Chern-Simons theory at level $n$ shown to be exactly equivalent to
the Laughlin theory at filling fraction $1/n$. Recently it is
commonly acceptable that the noncommutativity would occur in the
spacetime with the Planck scale and the NCFT~\cite{s4} would play an
important role in describing phenomena at planckian regimes.

The mathematical background for the NCFT is the noncommutative geometry~\cite{s5}.
In general, the spacetime noncommutativity can be distinguished
in accordance with the Hopf-algebraic classification by the three types,
{\em i.e.} the canonical, Lie-algebraic and quadratic
noncommutativity, respectively. Among them, the ${\kappa}$-deformed Minkowski spacetime~\cite{s6,s7}
as a specific case of the Lie-algebraic type
has recently been paid much attention because it is a natural candidate
for the spacetime based on which the Doubly Special Relativity~\cite{s8} has been established.
The ${\kappa}$-Minkowski spacetime is defined by the following commutation relations\footnote{Here the noncommutative
parameter ${\kappa}$ with the mass dimension is considered to be real and positive.} of the Lie-algebraic type,
\begin{equation}
[{\hat x}^0,{\hat x}^j]=\frac{i}{\kappa}{\hat x}^j, \qquad [{\hat x}^i,{\hat x}^j]=0, \qquad i,j=1,2,3.
\end{equation}
In addition, the vanishing momentum commutation relations taken from the ${\kappa}$-deformed Poincar$\acute{\rm e}$
algebra should be supplemented,
\begin{equation}
[{\hat p}_{\mu},{\hat p}_{\nu}]=0, \qquad {\mu},{\nu}=0,1,2,3.
\end{equation}
Therefore eqs.~(1) and (2), together with the cross relations between ${\hat x}^{\mu}$ and ${\hat p}_{\nu}$ that should
coincide with the Jacobi identity, constitute a complete algebra of the noncommutative phase space.

Although we focus in this paper on an alternative scheme (see the next section for details)
on dealing with noncommutativity of spacetimes,\footnote{The concept of
noncommutative fields~\cite{s9} is different from that of the noncommutative spacetime and is not involved
in this paper.} here we briefly recapitulate the traditional scheme for the sake of presentation of a big difference
between the two schemes.
In the latter the noncommutativity can be described by the way of Weyl operators or for the sake of practical
applications
by the way of normal functions with a suitable definition of star-products.
The relationship between the two ways is, as was stated in ref.~\cite{s4}, that the noncommutativity of spacetimes
may be encoded through ordinary products in the
noncommutative $C^{\ast}$-algebra of Weyl operators, or equivalently through the deformation of the product of the
commutative $C^{\ast}$-algebra of
functions to a noncommutative star-product. For instance, in the canonical noncommutative spacetime
the star-product is just the Moyal-product~\cite{s10}, while in the ${\kappa}$-deformed Minkowski spacetime
the star-product requires a more complicated formula~\cite{s11}. Some recent progress on star-products shows~\cite{s12}
that a {\em local} field theory in the ${\kappa}$-deformed Minkowski spacetime can be described by a {\em nonlocal}
relativistic field theory in the Minkowski spacetime.

The arrangement of this paper is as follows. In the next section, a
new approach quite different from the traditional one mentioned above is proposed for disposing the
noncommutativity of the
${\kappa}$-Minkowski spacetime, that is, a noncommutative extension of the Minkowski spacetime
is introduced. As a byproduct,
the approach makes it possible that a {\em local} field
theory in the ${\kappa}$-deformed Minkowski spacetime can be reduced under some postulation of operator linearization
to a {\em still local} relativistic field theory in the noncommutative extension of the
Minkowski spacetime. In section 3, the idea of the extended Minkowski spacetime is applied as an example
to chiral bosons, and the lagrangian theories of noncommutative chiral bosons are established in such a
framework of the Minkowski spacetime and then quantized consistently in accordance with Dirac's method.
Furthermore, in section 4 the self-duality of the lagrangian theories is analyzed
in terms of the parent action method.
Finally section 5 is devoted to the conclusion and perspective, in which a kind of temporal
fuzziness related to noncommutativity is discussed in detail.

\section{Noncommutative extension of Minkowski spacetime}
In general the so-called noncommutativity is closely related to a noncommutative spacetime like the
${\kappa}$-deformed Minkowski spacetime, and it has nothing to do with a commutative spacetime,
such as the Minkowski spacetime.
In this paper we encode enough information of noncommutativity of the ${\kappa}$-Minkowski spacetime
to a commutative spacetime, and propose a noncommutative extension of the Minkowski spacetime.
This extended Minkowski spacetime is as commutative as the Minkowski spacetime,
however, it contains noncommutativity in fact.
Our motive originates from a new point of view of disposing noncommutativity,
that is, whether the ${\kappa}$-deformed Minkowski spacetime can be dealt
with in some sense in the same way as the Minkowski spacetime.
This idea may be realized by connecting a commutative spacetime
with the ${\kappa}$-Minkowski spacetime and by introducing a well-defined proper time.
As a result, the
noncommutative field theories defined in the ${\kappa}$-deformed Minkowski spacetime may be investigated somehow
by the way of the ordinary (commutative) field theories in the noncommutative extension of the Minkowski spacetime
and thus the noncommutativity may be depicted within the framework of this commutative spacetime.
That is, we may provide from the point of view of noncommutativity a simplified treatment to
the noncommutativity of the ${\kappa}$-Minkowski spacetime,
which indeed unveils certain interesting features, such as fuzziness in the temporal dimension.

We start with an algebra of coordinate and momentum operators of a commutative spacetime,
which is nothing but the standard Heisenberg commutation
relations,\footnote{The speed of light $c$ and the Planck constant $\hbar$ are set to unit throughout this paper.}
\begin{equation}
[{\hat{\cal X}}^{\mu},{\hat{\cal X}}^{\nu}]=0, \qquad
[{\hat{\cal X}}^{\mu},{\hat{\cal P}}_{\nu}]=i{\delta}^{\mu}_{\nu}, \qquad
[{\hat{\cal P}}_{\mu},{\hat{\cal P}}_{\nu}]=0.
\end{equation}
The connection between the commutative spacetime and the ${\kappa}$-deformed Minkowski spacetime
should be found because it is the basis for us to give the noncommutative extension of the Minkowski spacetime.
In accordance with ref.~\cite{s13} the following relations may be chosen,
\begin{eqnarray}
{\hat x}^0 & = & {\hat{\cal X}}^0-\frac{1}{\kappa}[{\hat{\cal X}}^j,{\hat{\cal P}}_j]_{+}, \\
{\hat x}^i & = & {\hat{\cal X}}^i+A{\eta}^{ij}{\hat{\cal P}}_j\exp\left(\frac{2}{\kappa}{\hat{\cal P}}_0\right),
\end{eqnarray}
where $[{\hat{\cal O}}_1,{\hat{\cal O}}_2]_{+} \equiv \frac{1}{2}\left({\hat{\cal O}}_1{\hat{\cal O}}_2
+{\hat{\cal O}}_2{\hat{\cal O}}_1\right)$, ${\eta}^{{\mu}{\nu}}\equiv{\rm diag}(1,-1,-1,-1)$,
and $A$ is an arbitrary constant.
This choice is, though not unique, rational because we are able to guarantee eq.~(1) in terms of eqs.~(3), (4) and (5).

According to the Casimir operator of the $\kappa$-deformed
Poincar{$\acute{\rm e}$} algebra on the standard basis~\cite{s6},
\begin{equation}
\hat{{\mathcal{C}}_{1}} = \left(2\kappa\sinh\frac{\hat{p}_0}{2\kappa}\right)^2-{\hat{p}_i}^2,
\end{equation}
we supplement the relations of momentum operators between the commutative spacetime and the ${\kappa}$-Minkowski
spacetime as follows:
\begin{eqnarray}
\hat{p}_0 & = & 2\kappa\,{\sinh}^{-1}\frac{{\hat\mathcal{P}}_0}{{2\kappa}}, \\
{\hat{p}_i} & = & {\hat\mathcal{P}}_i,
\end{eqnarray}
which obviously conform to eq.~(2). Now eqs.~(3), (4), (5), (7) and (8) can solely determine the cross commutators
between ${\hat x}^{\mu}$ and ${\hat p}_{\nu}$,
\begin{equation}
[{\hat x}^i,{\hat p}_j]=i{\delta}^{i}_{j}, \qquad
[{\hat x}^0,{\hat p}_0]=i\left(\cosh\frac{\hat{p}_0}{{2\kappa}}\right)^{-1},\qquad
[{\hat x}^0,{\hat p}_i]=-\frac{i}{\kappa}{\hat p}_i,\qquad
[{\hat x}^i,{\hat p}_0]=0.
\end{equation}
Fortunately, the complete algebra of the noncommutative phase space composed of eqs.~(1), (2) and (9) indeed satisfies
the Jacobi identity, which is, from the point of view of consistency, crucial to the relationship,
{\em i.e.} eqs.~(4), (5), (7) and (8) established
between the commutative spacetime and the ${\kappa}$-Minkowski spacetime.
Moreover, such a relationship makes the Casimir operator have the usual formula as expected,
\begin{equation}
\hat{{\mathcal{C}}_{1}} =  {{\hat\mathcal{P}}_0}^2-{{\hat\mathcal{P}}_i}^2,
\end{equation}
which is consistent with the standard Heisenberg commutation relations (eq.~(3)).

When $\hat{p}_{\mu}$ take the usual forms,
\begin{eqnarray}
\hat{p}_0 &=& -i\frac{\partial}{{\partial}t},\\
\hat{p}_i &=& -i\frac{\partial}{{\partial}x^i},
\end{eqnarray}
which coincide with eq.~(2), the operator ${\hat\mathcal{P}}_0$ then reads
\begin{equation}
{\hat\mathcal{P}}_0=-i{2\kappa}\left(\sin\frac{1}{2\kappa}\frac{{\partial}}{{\partial}t}\right).
\end{equation}
Next we demonstrate the meaning of $t$ and $x^i$ defined by eqs.~(11) and (12).
For simplicity and without contradiction to eqs.~(11) and (12), we set $A=0$ in eq.~(5). Therefore,
the canonical coordinates $x^i$ that are the eigenvalues of operators ${\hat{\cal X}}^{i}$
may be regarded as the eigenvalues of operators ${\hat x}^i$,
while the eigenvalue of operator ${\hat x}^0$ does not match
the ordinary time variable\footnote{The eigenvalue of operator ${\hat x}^0$ does not match the $t$-parameter either,
but takes a more complicated formula
that should coincide with the algebra of the noncommutative phase space defined by eqs.~(1), (2) and (9).}
because operator ${\hat x}^0$,
different from operator ${\hat{\cal X}}^0$, does not satisfy the standard Heisenberg commutation relations.
As a result, for the ${\kappa}$-deformed Minkowski spacetime
we have a certain degree of freedom to introduce a parameter
which is of the temporal dimension. This parameter that is required to tend to the ordinary time variable
in the limit $\kappa \rightarrow +\infty$
is used to describe the evolution of systems. Here $t$, introduced through eq.~(11), is
dealt with for the sake of simplicity as such a parameter, rather than the eigenvalue of operator ${\hat x}^0$,
to describe the dynamical evolution of fields.
It obviously tends to the ordinary time variable in the limit $\kappa \rightarrow +\infty$
as well as the eigenvalue of operator ${\hat x}^0$, which guarantees the consistency of the choice of the parameter.

Now we introduce a proper time\footnote{This terminology is adopted here only because of its temporal dimension.
It has nothing to do with that of the
special relativity.} $\tau$ by defining the operator
\begin{equation}
{\hat\mathcal{P}}_0 \equiv -i\frac{{\partial}}{{\partial}\tau}.
\end{equation}
$\tau$ may be treated as the eigenvalue of operator ${\hat{\cal X}}^0$, which, together with eq.~(14),
is in agreement with the standard Heisenberg commutation relations. At the present stage,
no connection between $\tau$ and $t$ can be determined.
Here a linear realization or representation of operator ${\hat\mathcal{P}}_0$ is postulated, which gives rise to
a well-defined proper time\footnote{This form is a natural choice that corresponds to a kind of minimal extensions
of the Minkowski spacetime from the point of view of noncommutativity.
The idea of minimal extensions is basic and usually adopted in physics, in particular at the beginning stage to
establish a theory, such as the minimal supersymmetry.
When we introduce light-cone components in the coordinate
system related to $\tau$, this form indeed induces models of noncommutative chiral bosons with interesting
physical properties. For the details, see further discussions.} that satisfies the following differential equation
in accordance with eqs.~(13) and (14),
\begin{equation}
2\kappa\left(\sin\frac{1}{2\kappa}\frac{d}{dt}\right)\tau=1.
\end{equation}
Fortunately, this equation has an exact solution, $\tau=t+\sum_{n=-\infty}^{+\infty}c_{n}\exp(2{\kappa}n{\pi}t)$,
where $n$ is an integer and coefficients $c_{n}$ are arbitrary real constants. The consistency requires
that $\tau$ should be convergent and tend
to parameter $t$ in the limit $\kappa \rightarrow +\infty$, which guarantees that both $\tau$ and $t$ can regress to
the ordinary time variable. This requirement nevertheless
adds\footnote{For the usual notation of $t \geq 0$. When $t < 0$, the solution takes the form,
$\tau=t+\sum_{n=1}^{+\infty}c^{\prime}_{n}\exp(2{\kappa}n{\pi}t)$, where $c^{\prime}_{n}$ are constants
and the convergence requires $c^{\prime}_{n}=0$ for ${n} < 0$.}
the constraints, $c_{n}=0$ for $n \geq 1$. Therefore, the final form of the solution reads\footnote{An additional
constraint should be imposed upon the non-vanishing coefficients, $c_{-n}$ for $n \geq 0$,
if the initial value of $\tau$ is required to be finite, which is related
to the so-called temporal fuzziness that will be analyzed in detail in the last section of this paper.}
\begin{equation}
\tau=t+\sum_{n=0}^{+\infty}c_{-n}\exp(-2{\kappa}n{\pi}t).
\end{equation}

The noncommutative extension of the Minkowski spacetime spanned by $(\tau, x^i)$ coordinates is thus given,
in which the Casimir operator and the line element have the same forms as that in the Minkowski spacetime. However,
some information of noncommutativity has been encoded through the proper time to the framework,
which can be seen clearly when this kind of extended spacetimes is transformed into
$(t, x^i)$ coordinates. This is one thing with two sides. Eq.~(16) plays a crucial role in connecting
the two coordinate systems to each other.
The connection means that a noncommutative spacetime, {\em i.e.} the $\kappa$-Minkowski spacetime
may be represented under the postulation of the operator
linearization (eq.~(15)) by a commutative spacetime, {\em i.e.} the extended Minkowski spacetime,
but the price paid for such a simplified treatment is the
appearance of infinitely many unfixed coefficients in the commutative spacetime.
This would be understandable because
it is just these unfixed coefficients that reflect the information of noncommutativity in the
commutative spacetime. See the last section
of this paper for a detailed discussion. By making use of eq.~(16) we then reduce
the Casimir operator to be
\begin{eqnarray}
\hat{{\mathcal C}^{\prime}_1} &=&-\frac{{\partial}^2}{{\partial}{\tau}^2}+\frac{{\partial}}{{\partial}x^i}
\frac{{\partial}}{{\partial}x^i} \nonumber \\
&=&-\frac{1}{{\dot \tau}}\frac{{\partial}}{{\partial}t}\left(\frac{1}{{\dot \tau}}\frac{{\partial}}{{\partial}t}\right)
+\frac{{\partial}}{{\partial}x^i}\frac{{\partial}}{{\partial}x^i},
\end{eqnarray}
and give the line element
\begin{eqnarray}
ds^2&=&d{\tau}^2-(dx^i)^2 \nonumber \\
&=&{\dot \tau}^2dt^2-(dx^i)^2,
\end{eqnarray}
where ${\dot \tau}$ means ${d\tau}/{dt}$,
\begin{equation}
{\dot \tau}=1-2{\kappa}{\pi}\sum_{n=0}^{+\infty}nc_{-n}\exp(-2{\kappa}n{\pi}t).
\end{equation}

The extended Minkowski spacetime is, as expected, commutative,
which is a merit for us to construct field theory models in this framework.
That is, we simply consider the Lorentz
invariance in the extended Minkowski spacetime, and the constructed models contain noncommutative effects naturally.
In mathematics, there exists a specific map (see eqs.~(4), (5), (7) and (8)) from the ${\kappa}$-Minkowski spacetime to
the noncommutative extension of the Minkowski spacetime.
The Lorentz invariance in the extended Minkowski spacetime reflects in fact some invariance
in the $\kappa$-Minkowski spacetime that seems an unknown symmetry up to now.\footnote{See, {\em e.g.}
ref.~\cite{s14} and the references therein.}
In this way we might have circumvented a relatively complicated procedure for searching for models that should possess
such an unknown symmetry.

The line element gives the fact that the noncommutative extension of the Minkowski spacetime
connects with a special flat spacetime corresponding to a twisted $t$-parameter.\footnote{The curvature of this spacetime equals to zero.
The author would like to thank the anonymous referee for pointing it out.}
In consequence we may say that
the ${\kappa}$-Minkowski spacetime, the source of the extended Minkowski spacetime,
is somehow reduced to the flat spacetime whose metric is given by
\begin{equation}
g_{00}={\dot \tau}^2=\left[1-2{\kappa}{\pi}\sum_{n=0}^{+\infty}nc_{-n}\exp(-2{\kappa}n{\pi}t)\right]^2,
\qquad g_{11}=g_{22}=g_{33}=-1.
\end{equation}

Before the end of this section we emphasize that the Casimir operator defined by eqs.~(6), (11) and (12)
with infinite order in $t$-parameter derivative has now been reduced\footnote{Here it only means that
$\hat{{\mathcal C}_1}$ is reduced to
$\hat{{\mathcal C}^{\prime}_1}$, while the inverse procedure is never implied in this paper.}
to the formula defined by eq.~(17)
with second order in $t$-parameter derivative under the postulation of the operator linearization (eq.~(15)).
This realization or representation of the Casimir operator (eq.~(17))
fulfills in a way the aim that some information of noncommutativity can be encoded to a commutative spacetime.
The feature of the extended Minkowski spacetime spanned by $(t, x^i)$ is that the evolution parameter is twisted
while the spaces are still flat (see, {\em e.g.} eq.~(18)),
which coincides with~\cite{s6} the characters of the
${\kappa}$-Minkowski spacetime, {\em i.e.} with a ``quantum time'' and a three-dimensional euclidean
space. This implies that the extended Minkowski spacetime {\em does}
contain enough information of noncommutativity that is able to reflect the characteristic of the
${\kappa}$-Minkowski spacetime although it does not recover the whole information of noncommutativity
due to the postulation of the operator linearization.

\section{Noncommutative chiral bosons}
We deviate from the discussion of noncommutativity temporarily and give a brief introduction of chiral bosons
in the Minkowski spacetime.
The main reason that chiral bosons\footnote{In general, one should mention chiral {\em p}-forms that include
chiral bosons as
the $p=0$ case. A chiral 0-form in the (1+1)-dimensional Minkowski spacetime is usually called a chiral boson
which describes a left- or right-moving boson in one spatial dimension.}
have received much attention
is that they appear in various theoretical models that relate to superstring theories, and
reflect especially the existence of a variety of important dualities that connect these theories
among one another. One has to envisage two basic problems in a lagrangian description of
chiral bosons: the first is the consistent quantization and the second is the harmonic combination
of the manifest duality and Lorentz covariance, since the equation of motion of a chiral boson,
{\em i.e.} the self-duality condition is first order with respect to the derivatives of space and time.
In order to solve these problems, various types of formulations of chiral bosons,
each of which possesses its own advantages, have been proposed~\cite{s15,s16,s17}.
It is remarkable that these models of chiral bosons have close relationships among one another, especially
various dualities that have been
demonstrated in detail from the points of view of both the configuration~\cite{s18,s19} and momentum~\cite{s20} spaces.

In this section we mainly propose noncommutative chiral bosons\footnote{We
can also consider chiral {\em p}-forms ($p\geq 1$) and their noncommutative generalizations.
This is one of the further topics that will probably be reported elsewhere soon.} in the noncommutative
extension of the Minkowski spacetime.
The method is as follows: a lagrangian of noncommutative chiral bosons is given
simply by the requirement of the Lorentz invariance in the extended Minkowski spacetime spanned by the
coordinates $(\tau, x)$, and through the coordinate transformation eq.~(16),
it is then converted into its $(t, x)$-coordinate formulation with explicit noncommutativity.
As a result, we establish the lagrangian theory of noncommutative chiral bosons in the extended framework of
the Minkowski spacetime. Alternatively,
we can also construct a lagrangian theory directly in the  $(t, x)$-coordinate framework in terms of the metric eq.~(20).
It should be noted that for a certain formulation of chiral bosons both procedures give rise to the noncommutative
generalizations that have the same physical spectrum.

Let us begin from the light-cone coordinates and their derivatives defined in the $(1+1)$-dimensional noncommutative
extension of the Minkowski spacetime, respectively, as follows:
\begin{eqnarray}
X^{\pm} & \equiv & \frac{1}{\sqrt 2}\left(\pm{\tau}+x\right),\\
D_{\pm} & \equiv & \frac{1}{\sqrt 2}\left(\pm\frac{{\partial}}{{\partial}\tau}+\frac{{\partial}}{{\partial}x}\right).
\end{eqnarray}
It is obvious that they satisfy $D_{\pm}X^{\pm}=1$ and $D_{\pm}X^{\mp}=0$. In the spacetime spanned by the
coordinates $(\tau, x)$ the equation of motion for a noncommutative chiral boson takes the usual form
\begin{equation}
D_{\mp}{\phi}=0,
\end{equation}
and its solution thus reads
\begin{equation}
{\phi}={\phi}(X^{\pm}),
\end{equation}
where the upper sign corresponds to the left-moving while the lower the right-moving.
Through the coordinate transformation we convert the equation of motion  eq.~(23) into its corresponding formulation
in the coordinate system spanned by $(t, x)$,
\begin{equation}
\mp\frac{1}{{\dot \tau}}\frac{{\partial}\phi}{{\partial}t}+\frac{{\partial}\phi}{{\partial}x}=0.
\end{equation}
The solution takes the same form as eq.~(24) and can easily be expressed by the $(t,x)$ coordinates through the
well-defined $X^{\pm}$,
\begin{equation}
{\phi}={\phi}\left(\frac{1}{\sqrt 2}\left[\pm t+x \pm \sum_{n=0}^{+\infty}c_{-n}
\exp\left(-2{\kappa}n{\pi}t\right)\right]\right).
\end{equation}
Consequently, the equation of motion and its solution comprise in a natural way the noncommutative effects
related to the finite noncommutative parameter ${\kappa}$.
Incidentally, they become their ordinary forms correspondent to the Minkowski spacetime in the limit
$\kappa \rightarrow +\infty$.

The following task is straightforward, that is, to construct in the extended framework of the Minkowski
spacetime such a lagrangian that yields the equation of motion (eq.~(25)) for noncommutative chiral bosons
by the method mentioned in the second paragraph of this section.
Three typical formulations of chiral bosons are investigated
as a primary application in this section,
{\em i.e.} the non-manifestly Lorentz covariant version~\cite{s15} and manifestly
Lorentz covariant versions with the linear self-duality constraint~\cite{s16} and with the quadratic one~\cite{s17},
respectively.

\subsection{The non-manifestly Lorentz covariant formulation}
It is non-manifestly Lorentz covariant but indeed Lorentz invariant~\cite{s15}.
Simply considering the Lorentz invariance in the extended Minkowski spacetime $(\tau, x)$, we give the action
\begin{equation}
S_{1}=\int d{\tau}dx\left[\frac{{\partial}\phi}{{\partial}\tau}\frac{{\partial}\phi}{{\partial}x}
-\left(\frac{{\partial}\phi}{{\partial}x}\right)^2\right],
\end{equation}
which is nothing but the formulation of Floreanini and Jackiw's left-moving chiral bosons
if $\tau$ is replaced by the ordinary time. After making the coordinate transformation, we
obtain the action written in terms of the coordinates $(t,x)$,
\begin{equation}
S_{1}=\int dtdx\sqrt{-g}\left[\frac{1}{{\dot \tau}}\frac{{\partial}\phi}{{\partial}t}
\frac{{\partial}\phi}{{\partial}x}-\left(\frac{{\partial}\phi}{{\partial}x}\right)^2\right],
\end{equation}
where $\sqrt{-g}$ is the Jacobian and also the nontrivial measure of the flat spacetime eq.~(20) connected with the
${\kappa}$-Minkowski spacetime,
\begin{equation}
\sqrt{-g}=\vert{\dot \tau}\vert.
\end{equation}
Therefore the lagrangian takes the form
\begin{equation}
{\cal L}_{1}=\pm{\dot \phi}{\phi}^{\prime} -\sqrt{-g}{{\phi}^{\prime}}^2,
\end{equation}
where a dot and a prime stand for derivatives with respect to time\footnote{For the sake of convenience
in description, {\em time}, here different from the ordinary time variable, stands only for {\em t-parameter}
in the three subsections of section 3.}
$t$ and space $x$, respectively.
A plus and minus sign appears in front of the first term due to the ratio $\sqrt{-g}/{\dot \tau}$, and the choice
depends on either ${\dot \tau} > 0$ or ${\dot \tau} < 0$. However, this does not cause any
ambiguity.\footnote{The same result can be obtained
if we start with the action for right-moving chiral bosons, which is also available in subsections 3.2 and 3.3.}
By making use of Dirac's quantization~\cite{s21} we can prove
that the lagrangian only describes a left-moving noncommutative chiral boson
in the extended framework of the Minkowski spacetime, which is independent of whether ${\dot \tau} > 0$
or ${\dot \tau} < 0$, and that the similar case
also happens in the linear and quadratic self-duality constraint formulations of chiral bosons.
Incidentally, this feature does not exist in the case of ordinary (commutative) chiral bosons
due to the triviality $\sqrt{-g}={\dot \tau}=1$ in the limit $\kappa \rightarrow +\infty$.

In terms of Dirac's quantization~\cite{s21} we can verify that the lagrangian ${\cal L}_{1}$ indeed describes a
noncommutative chiral boson
which satisfies the equation of motion eq.~(25) with the choice of the upper sign correspondent to the left-handed
chirality. To this end, at first define the momentum conjugate to $\phi$,
\[
{\pi}_{\phi} \equiv \frac{{\partial}{\cal L}_{1}}{{\partial}{\dot \phi}}=\pm{\phi}^{\prime},
\]
and then give the hamiltoian through the Legendre transformation,
\[
{\cal H}_{1}={\pi}_{\phi}{\dot \phi}-{\cal L}_{1}=\sqrt{-g}{{\phi}^{\prime}}^2.
\]
Note that this hamiltoian explicitly contains
time\footnote{The Dirac quantization is shown to be available in constrained systems whose hamiltoians contain time
explicitly. For instance, see ref.~\cite{s22}.} and that
it is positive definite as its counterpart in the Minkowski spacetime. The definition of momenta actually yields one
primary constraint
\[
{\Omega}(x)\equiv {\pi}_{\phi} \mp {\phi}^{\prime} \approx 0,
\]
where ``$\approx$'' stands for Dirac's weak equality.
Because of no further constraints and of the non-vanishing equal-time Poisson bracket,
\[
\{{\Omega}(x),{\Omega}(y)\}_{PB}=\mp 2{\partial}_{x}{\delta}(x-y),
\]
this constraint itself constitutes a second-class set. With the inverse of the Poisson bracket,
\[
\{{\Omega}(x),{\Omega}(y)\}^{-1}_{PB}=\mp \frac{1}{2}{\varepsilon}(x-y),
\]
where ${\varepsilon}(x)$ is the step function with the property
$d{\varepsilon}(x)/dx=\delta(x)$, we calculate the equal-time Dirac brackets:
\begin{eqnarray*}
\{{\phi}(x),{\phi}(y)\}_{DB}&=&\mp \frac{1}{2}{\varepsilon}(x-y),\nonumber \\
\{{\phi}(x),{\pi}_{\phi}(y)\}_{DB}&=&\frac{1}{2}\delta(x-y),\nonumber \\
\{{\pi}_{\phi}(x),{\pi}_{\phi}(y)\}_{DB}&=&\pm \frac{1}{2}{\partial}_{x}\delta(x-y).
\end{eqnarray*}
In the sense of Dirac brackets
weak constraints become strong conditions. As a consequence, we write the reduced hamiltonian
\[
{\cal H}^{r}_{1}=\sqrt{-g}{{\phi}^{\prime}}^2=\sqrt{-g}{\pi}_{\phi}^2
=\frac{1}{2}\sqrt{-g}\left({{\phi}^{\prime}}^2+{\pi}_{\phi}^2\right),
\]
and derive from the canonical hamiltonian equation
\[
{\dot \phi}=\int dy \{{\phi}(x),{\cal H}^{r}_{1}(y)\}_{DB},
\]
the equation of motion for the noncommutative chiral boson,
\[
{\dot \phi}=\pm \sqrt{-g}{\phi}^{\prime}={\dot \tau}{\phi}^{\prime},
\]
which is nothing but eq.~(25) with the upper sign corresponding to the left-handed chirality.

\subsection{The manifestly Lorentz covariant formulation with linear self-duality constraint}
In this formulation the self-duality constraint is imposed upon a massless real scalar field~\cite{s16}.
Although it has some defects~\cite{s23}, the linear formulation strictly describes a chiral boson from the point of
view of equations of motion
at both the classical and quantum levels. Its generalization in the canonical noncommutative spacetime has been
studied in detail, and in particular a kind of fuzziness on the left- and right-handed chiralities in the spatial
dimension has been noticed~\cite{s22}.
Under the requirement of the Lorentz invariance in the extended Minkowski spacetime, we can write the action in a
straightforward way,
\begin{equation}
S_{2}=\int d{\tau}dx\left(-D_{+}{\phi}D_{-}{\phi}-\sqrt{2}{\lambda}_{+}D_{-}{\phi}\right),
\end{equation}
where ${\lambda}_{+} \equiv \frac{1}{\sqrt{2}}({\lambda}_{0}+{\lambda}_{1})$, a light-cone component of the vector
field ${\lambda}_{\mu}$, ${\mu}=0,1$, introduced
as a Lagrange multiplier.
In terms of the method adopted in subsection 3.1, we then convert it into its formulation in the
framework spanned by the coordinates $(t, x)$,
\begin{equation}
S_{2}=\int dtdx\sqrt{-g}\left[\frac{1}{2{\dot \tau}^2}\left(\frac{{\partial}\phi}{{\partial}t}\right)^2-\frac{1}{2}
\left(\frac{{\partial}\phi}{{\partial}x}\right)^2+{\lambda}_{+}
\left(\frac{1}{{\dot \tau}}\frac{{\partial}\phi}{{\partial}t}-\frac{{\partial}\phi}{{\partial}x}\right)\right],
\end{equation}
from which the lagrangian reads
\begin{equation}
{\cal L}_{2}=\frac{1}{2\sqrt{-g}}\left[{\dot \phi}^2-\left(\sqrt{-g}{\phi}^{\prime}\right)^2\right]
+{\lambda}_{+}\left(\pm{\dot \phi}-\sqrt{-g}{\phi}^{\prime}\right),
\end{equation}
where a plus and minus sign also exists in front of ${\dot \phi}$ as explained in subsection 3.1.

As was done in the above subsection, we make the hamiltonian analysis by using Dirac's method and prove that
${\cal L}_{2}$ describes, as expected,
a noncommutative chiral boson with the left-handed chirality. Let us define momenta conjugate to ${\phi}$ and
${\lambda}_{+}$, respectively,
\begin{eqnarray}
{\pi}_{\phi} & \equiv & \frac{{\partial}{\cal L}_{2}}{{\partial}{\dot \phi}}
= \frac{1}{\sqrt{-g}}{\dot \phi} \pm {\lambda}_{+},
\nonumber \\
{\pi}_{{\lambda}_{+}} & \equiv & \frac{{\partial}{\cal L}_{2}}{{\partial}{\dot {\lambda}_{+}}}  \approx  0.\nonumber
\end{eqnarray}
The latter gives in fact one primary constraint
\[
{\Omega}_{1}(x)\equiv {\pi}_{{\lambda}_{+}} \approx 0.
\]
Making the Legendre transformation, we get the canonical hamiltonian
\begin{eqnarray*}
{\cal H}_{2} & =& {\pi}_{\phi}{\dot \phi}+{\pi}_{{\lambda}_{+}}{\dot {\lambda}_{+}}-{\cal L}_{2}\nonumber \\
& =& \sqrt{-g}\left[\frac{1}{2}\left({\pi}_{\phi}^2
+{{\phi}^{\prime}}^2\right)+{\lambda}_{+}\left(\mp{\pi}_{\phi}+{\phi}^{\prime}\right)
+\frac{1}{2}{\lambda}_{+}^2\right].
\end{eqnarray*}
As a basic consistency requirement in dynamics of constrained systems, ${\Omega}_{1}(x)$ should be preserved in time,
which yields one secondary constraint
\[
{\Omega}_{2}(x)\equiv \pm {\pi}_{\phi}-{\phi}^{\prime}-{{\lambda}_{+}} \approx 0.
\]
Because the preservation of
${\Omega}_{2}(x)$ does not give further constraints, the constraint
set consists of ${\Omega}_{1}(x)$ and ${\Omega}_{2}(x)$. With the matrix elements of equal-time Poisson
brackets of the constraints,
\begin{eqnarray}
& & C_{11}(x,y)=0,\nonumber \\
& & C_{12}(x,y)=-C_{21}(x,y)=\delta(x-y),\nonumber \\
& & C_{22}(x,y)=\mp 2{\partial}_{x}{\delta}(x-y),\nonumber
\end{eqnarray}
we can easily deduce their inverse elements
\begin{eqnarray}
& & C^{-1}_{11}(x,y)=\mp 2{\partial}_{x}{\delta}(x-y),\nonumber \\
& & C^{-1}_{12}(x,y)=-C^{-1}_{21}(x,y)=-\delta(x-y),\nonumber \\
& & C^{-1}_{22}(x,y)=0,\nonumber
\end{eqnarray}
and then compute the non-vanishing equal-time Dirac brackets
\begin{eqnarray}
\{{\phi}(x),{\pi}_{\phi}(y)\}_{DB} &=& \delta(x-y),\nonumber \\
\{{\phi}(x),{\lambda}_{+}(y)\}_{DB} &=& \pm \delta(x-y),\nonumber \\
\{{\pi}_{\phi}(x),{\lambda}_{+}(y)\}_{DB} &=& -{\partial}_{x}{\delta}(x-y),\nonumber \\
\{{\lambda}_{+}(x),{\lambda}_{+}(y)\}_{DB} &=& \mp 2{\partial}_{x}{\delta}(x-y).\nonumber
\end{eqnarray}
After making Dirac's weak constraints
be strong conditions, we obtain the reduced hamiltonian in terms of independent phase space variables,
\[
{\cal H}^{r}_{2}=\pm \sqrt{-g}{\pi}_{\phi}{{\phi}^{\prime}}.
\]
Note that the linear self-duality model
of chiral bosons in the ordinary spacetime~\cite{s16} is intrinsically non-positive
definite, the non-positive definition here should not be induced by its generalization in the extended Minkowski
spacetime but emerges from the original (commutative) formulation. From the canonical hamiltonian equation,
\[
{\dot \phi}=\int dy \{{\phi}(x),{\cal H}^{r}_{2}(y)\}_{DB},
\]
we thus arrive at the expected equation of motion
\[
{\dot \phi}=\pm \sqrt{-g}{\phi}^{\prime}={\dot \tau}{\phi}^{\prime}.
\]

\subsection{The manifestly Lorentz covariant formulation with quadratic self-duality constraint}
In this formulation the square of the self-duality constraint, instead of the self-duality itself,
is imposed upon a massless real scalar field~\cite{s17}.
In accordance with the Lorentz invariance we firstly write the following action in the extended Minkowski
spacetime related to the coordinates $(\tau, x)$,
\begin{equation}
S_{3}=\int d{\tau}dx\left[-D_{+}{\phi}D_{-}{\phi}-{\lambda}_{++}\left(D_{-}{\phi}\right)^2\right],
\end{equation}
where ${\lambda}_{++} \equiv \frac{1}{2}({\lambda}_{00}+{\lambda}_{01}+{\lambda}_{10}+{\lambda}_{11})$, a light-cone
component of the tensor field ${\lambda}_{{\mu}{\nu}}$, ${\mu}, {\nu}=0,1$, introduced as a Lagrange multiplier,
and then rewrite it in the $(t, x)$-coordinate framework in terms of eq.~(16),
\begin{equation}
S_{3}=\int dtdx\sqrt{-g}\left[\frac{1}{2{\dot \tau}^2}\left(\frac{{\partial}\phi}{{\partial}t}\right)^2-\frac{1}{2}
\left(\frac{{\partial}\phi}{{\partial}x}\right)^2-\frac{1}{2}{\lambda}_{++}
\left(\frac{1}{{\dot \tau}}\frac{{\partial}\phi}{{\partial}t}-\frac{{\partial}\phi}{{\partial}x}\right)^2\right],
\end{equation}
which yields at last the lagrangian
\begin{equation}
{\cal L}_{3}=\frac{1}{2\sqrt{-g}}\left[{\dot \phi}^2-\left(\sqrt{-g}{\phi}^{\prime}\right)^2\right]
-\frac{1}{2\sqrt{-g}}{\lambda}_{++}\left(\pm{\dot \phi}-\sqrt{-g}{\phi}^{\prime}\right)^2,
\end{equation}
where a plus and minus sign emerges in front of ${\dot \phi}$ once again as occurred in subsections 3.1 and 3.2.

Briefly repeating the procedure gone through in the above two
subsections, we can make the conclusion that ${\cal L}_{3}$ also
describes a noncommutative chiral boson with the left-handed chirality in the
extended Minkowski spacetime.
At first, through introducing canonical momenta conjugate to
${\phi}$ and ${\lambda}_{++}$, respectively,
\begin{eqnarray}
{\pi}_{\phi} & \equiv & \frac{{\partial}{\cal L}_{3}}{{\partial}{\dot \phi}}=\frac{1}{\sqrt{-g}}
\left(1-{\lambda}_{++}\right){\dot\phi} \pm {\lambda}_{++}{\phi}^{\prime},\nonumber \\
{\pi}_{{\lambda}_{++}} & \equiv & \frac{{\partial}{\cal L}_{3}}{{\partial}{\dot {\lambda}_{++}}} \approx 0,\nonumber
\end{eqnarray}
we get a primary constraint
\[
{\Omega}_{1}(x)\equiv {\pi}_{{\lambda}_{++}} \approx 0,
\]
and derive the canonical hamiltonian in terms of the Legendre transformation,
\[
{\cal H}_{3}={\pi}_{\phi}{\dot \phi}+{\pi}_{{\lambda}_{++}}{\dot{\lambda}_{++}}-{\cal L}_{3}
=\frac{\sqrt{-g}}{1-{\lambda}_{++}}\left(\frac{1}{2}{\pi}_{\phi}^2 \mp
{\lambda}_{++}{\pi}_{\phi}{\phi}^{\prime}+\frac{1}{2}{{\phi}^{\prime}}^2\right).
\]
The consistency of time evolution of ${\Omega}_{1}(x)$ then gives rise to one secondary constraint
\[
{\tilde{\Omega}}_{2}(x) \equiv \left({\pi}_{\phi} \mp {\phi}^{\prime}\right)^2 \approx 0,
\]
which is first-class and no longer induces further constraints. At this
stage, we may replace ${\tilde{\Omega}}_{2}(x)$ by its linearized version,
\[
{\Omega}_{2}(x) \equiv {\pi}_{\phi} \mp {\phi}^{\prime} \approx 0,
\]
which is second-class, as was dealt with~\cite{s24} to
the ordinary quadratic self-duality constraint formulation in the
Minkowski spacetime under the consideration of the classical
equivalence between the two constraints. According to Dirac's
method, a gauge fixing condition
\[
{\chi}(x)\equiv {\lambda}_{++}(x)-F(x)\approx 0
\]
with $F(x)$ an arbitrary function in the extended framework of the Minkowski spacetime,
should be added, and thus the constraint set, $({\chi}(x),
{\Omega}_{1}(x), {\Omega}_{2}(x))$, becomes second-class. The
remainder of the canonical analysis can be followed straightforwardly
and the results, such as the non-vanishing equal-time Dirac
brackets, the reduced hamiltonian, and the equation of motion, are
exactly same as that obtained in subsection 3.1. As a consequence, we
verify that the quadratic formulation can be reduced to the
noncommutative generalization of Floreanini and Jackiw's chiral
bosons, which reveals the connection between the two formulations in
the noncommutative extension of the Minkowski spacetime.

\section{Self-duality of noncommutative chiral bosons}
Based on the important role played by the duality and/or self-duality in the ordinary chiral
$p$-forms and the related theories~\cite{s18, s19, s20}, it is curious to
argue whether such a symmetry maintains or not in their various noncommutative generalizations.
It was revealed~\cite{s22} that the self-duality
is not preserved when the chiral boson action with the linear chirality constraint~\cite{s16} is generalized from
the Minkowski spacetime to the canonical noncommutative spacetime. However,
it will be verified in the following context that this kind of symmetries still exists in the
generalizations performed in the noncommutative extension of the Minkowski spacetime.
The situation implies somehow a complex relationship between duality/self-duality and noncommutativity,
which may be worth notice.

In this section a systematic approach, known as the parent action method~\cite{s25}, is used to investigate
the self-duality of the actions eqs.~(28), (32) and (35).
The approach, from the Legendre transformation and for the foundation of equivalence among theories
at the level of actions instead of equations of motion,
can be summarized\footnote{As to the original proposal and recent development of the parent action method, see,
for instance, the references cited by ref.~\cite{s25}.}
briefly as follows:
(i) to introduce auxiliary fields and then to construct a parent or master
action based on a source action, and (ii) to make variation of the parent
action with respect to each auxiliary field, to solve one auxiliary field
in terms of other fields and then to substitute the solution into the parent
action. Through making variations with respect to different auxiliary fields,
we can obtain different forms of an action. These forms are, of
course, equivalent classically, and the relation between them is usually
referred to duality. If the resulting forms are same,
their relation is called self-duality.

\subsection{Model in subsection 3.1}
According to the parent action approach~\cite{s25}, we introduce
two auxiliary vector fields $G^{\mu}$ and $F_{\mu}$,
and write down\footnote{For simplicity, we only consider case ${\dot \tau} > 0$
in the discussion of self-duality in this section. As to case ${\dot \tau} < 0$,
the same result will be deduced.}
the parent action correspondent to $S_{1}$, {\em i.e.} eq.~(28),
\begin{equation}
S_{1}^{{\rm p}}=\int dt dx\left[F_0F_1-\sqrt{-g}{F_1}^2+G^{\mu}\left(F_{\mu}-{\partial}_{\mu}\phi\right)\right].
\end{equation}
Now varying eq.~(37) with respect to $G^{\mu}$ gives
$F_{\mu}={\partial}_{\mu}\phi$, together with which eq.~(37) regresses to the
action eq.~(28). This shows the classical equivalence between the parent action $S_{1}^{{\rm p}}$ and
its source action $S_{1}$. However, varying eq.~(37) with respect to $F_{\mu}$
leads to the expression of $F_{\mu}$ in terms of $G^{\mu}$:
\begin{eqnarray}
F_0 &=& -2\sqrt{-g}G^0-G^1,\nonumber \\
F_1 &=& -G^0.
\end{eqnarray}
Substituting eq.~(38) into eq.~(37),
we obtain a kind of dual versions for the action $S_{1}$,
\begin{equation}
{\tilde S}_{1}^{{\rm dual}}=\int dt dx\left[-G^0G^1-\sqrt{-g}\left({G^0}\right)^2+\phi{\partial}_{\mu}G^{\mu}\right],
\end{equation}
where $\phi$ is dealt with at present as a Lagrangian multiplier.
Further varying eq.~(39) with respect to $\phi$
gives ${\partial}_{\mu}G^{\mu}=0$, whose solution takes the form
\begin{equation}
G^{\mu}={\epsilon}^{{\mu}{\nu}}{\partial}_{\nu}\varphi,
\end{equation}
where ${\epsilon}^{00}={\epsilon}^{11}=0$, ${\epsilon}^{01}=-{\epsilon}^{10}=+1$,
and $\varphi$ is a scalar filed that is in general different from $\phi$.
Therefore the dual version eq.~(39) can be reduced to its simpler formula described only by $\varphi$,
\begin{equation}
S_{1}^{{\rm dual}}=\int dt dx\left[{\dot \varphi}{\varphi}^{\prime} -\sqrt{-g}{{\varphi}^{\prime}}^2\right].
\end{equation}
It has the same form as eq.~(28) just with the replacement of $\phi$ by $\varphi$, that is,
eq.~(28) or eq.~(41) is self-dual with respect to the chiral boson field.

In accordance with the illustration
utilized for duality in ref.~\cite{s25}, the self-duality of $S_{1}$ and $S_{1}^{{\rm dual}}$ may be
illustrated by Fig.~1.
\vspace{3mm}
\begin{figure}[htb]
\begin{center}
\setlength{\unitlength}{.7mm}
\begin{picture}(100,50)(0,10)
\put(46,60){$S_{1}^{{\rm p}}$}
\put(46,57){\vector(-2,-3){23}}\put(54,57){\vector(2,-3){23}}
\put(22,40){${\delta G^{\mu}}$}\put(68,40){$\delta F_{\mu}+\delta
\phi$} \put(18,17){${S_{1}}$}\put(78,17){${S_{1}^{{\rm dual}}}$}
\put(20,10){\vector(0,1){5}}\put(80,10){\vector(0,1){5}}
\put(20,10){\line(1,0){16}}\put(64,10){\line(1,0){16}}
\put(38,9){\bf{self-dual}}
\end{picture}
\caption{\small Schematic relation of the actions: the parent action
$S_{1}^{{\rm p}}$ shows that $S_{1}$ and $S_{1}^{{\rm dual}}$ have
self-duality with respect to the chiral boson field.}
\end{center}
\end{figure}
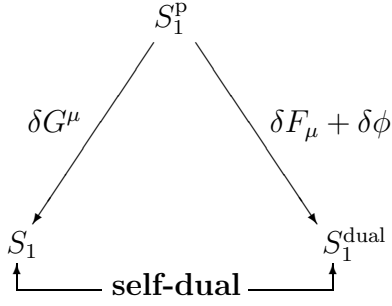

\subsection{Model in subsection 3.2}
Following the procedure gone through in the above subsection, we can verify that $S_{2}$ (eq.~(32))
is self-dual with respect to the chiral boson field.
To this end, let us at first regard $S_{2}$ as a source action and
give its corresponding parent action
\begin{equation}
S_{2}^{{\rm p}}=\int dt dx\left\{\frac{1}{2\sqrt{-g}}\left[{F_0}^2-\left(\sqrt{-g}F_1\right)^2\right]
+{\lambda}_{+}\left({F_0}-\sqrt{-g}F_1\right)+G^{\mu}\left(F_{\mu}-{\partial}_{\mu}\phi\right)\right\},
\end{equation}
where $G^{\mu}$ and $F_{\mu}$ stand for auxiliary vector fields as before.
Next, through making the variation of eq.~(42) with respect to $G^{\mu}$ we simply
obtain $F_{\mu}={\partial}_{\mu}\phi$, with which eq.~(42) becomes the source action eq.~(32). This implies that
$S_{2}^{{\rm p}}$ is classically equivalent to $S_{2}$. Thirdly, by making the variation
of eq.~(42) with respect to $F_{\mu}$ we have
\begin{eqnarray}
F_0 &=& -\sqrt{-g}\left(G^0+{\lambda}_{+}\right),\nonumber \\
F_1 &=& \frac{1}{\sqrt{-g}}\left(G^1-\sqrt{-g}{\lambda}_{+}\right).
\end{eqnarray}
After substituting eq.~(43) into eq.~(42) we then deduce a dual version of $S_{2}$,
\begin{equation}
{\tilde S}_{2}^{{\rm dual}}=\int dt dx\left\{-\frac{1}{2{\sqrt{-g}}}\left[\left({\sqrt{-g}}G^0\right)^2-
\left(G^1\right)^2\right]-{\lambda}_{+}\left(\sqrt{-g}{G^0}+G^1\right)+\phi{\partial}_{\mu}G^{\mu}\right\}.
\end{equation}
At last, varying eq.~(44) with respect to $\phi$ which is now dealt with as a Lagrangian multiplier, we
derive eq.~(40) again and therefore simplify the above dual action to be
\begin{equation}
S_{2}^{{\rm dual}}=\int dt dx\left\{\frac{1}{2\sqrt{-g}}\left[{\dot \varphi}^2
-\left(\sqrt{-g}{\varphi}^{\prime}\right)^2\right]
+{\lambda}_{+}\left({\dot \varphi}-\sqrt{-g}{\varphi}^{\prime}\right)\right\},
\end{equation}
which is nothing but eq.~(32) if $\varphi$ is replaced by $\phi$. The illustration of
self-duality of $S_{2}$ and $S_{2}^{{\rm dual}}$ can be seen in Fig.~2.
\vspace{3mm}
\begin{figure}[htb]
\begin{center}
\setlength{\unitlength}{.7mm}
\begin{picture}(100,50)(0,10)
\put(46,60){$S_{2}^{{\rm p}}$}
\put(46,57){\vector(-2,-3){23}}\put(54,57){\vector(2,-3){23}}
\put(22,40){${\delta G^{\mu}}$}\put(68,40){$\delta F_{\mu}+\delta
\phi$} \put(18,17){${S_{2}}$}\put(78,17){${S_{2}^{{\rm dual}}}$}
\put(20,10){\vector(0,1){5}}\put(80,10){\vector(0,1){5}}
\put(20,10){\line(1,0){16}}\put(64,10){\line(1,0){16}}
\put(38,9){\bf{self-dual}}
\end{picture}
\caption{\small Schematic relation of the actions: the parent action
$S_{2}^{{\rm p}}$ shows that $S_{2}$ and $S_{2}^{{\rm dual}}$ have
self-duality with respect to the chiral boson field.}
\end{center}
\end{figure}
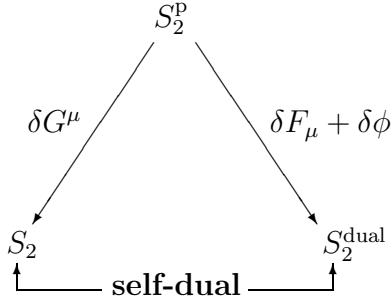

\subsection{Model in subsection 3.3}
Simply repeating the procedure followed in the above two subsections, we can easily show the self-duality of $S_{3}$
(eq.~(35)) with respect to the chiral boson field $\phi$. The parent action, correspondent to $S_{3}$ as a source,
takes the form
\begin{equation}
S_{3}^{{\rm p}}=\int dt dx\left\{
\frac{1}{2\sqrt{-g}}\left[{F_0}^2-\left(\sqrt{-g}F_1\right)^2\right]
-\frac{1}{2\sqrt{-g}}{\lambda}_{++}\left({F_0}-\sqrt{-g}F_1\right)^2 
+G^{\mu}\left(F_{\mu}-{\partial}_{\mu}\phi\right)\right\},
\end{equation}
where $G^{\mu}$ and $F_{\mu}$ are auxiliary vector fields introduced. Varying the parent action
with respect to $G^{\mu}$
gives $F_{\mu}={\partial}_{\mu}\phi$, which provides nothing new but just the classical equivalence between
$S_{3}^{{\rm p}}$ and $S_{3}$. However, varying it with respect to $F_{\mu}$ instead leads to the useful relations
between the two auxiliary fields,
\begin{eqnarray}
F_0 &=& -\sqrt{-g}\left(1+{\lambda}_{++}\right)G^0-{\lambda}_{++}G^1,\nonumber \\
F_1 &=& -{\lambda}_{++}G^0+\frac{1}{\sqrt{-g}}\left(1-{\lambda}_{++}\right)G^1.
\end{eqnarray}
Substituting the above equation into the parent action, we derive a dual action described by $G^{\mu}$ and $\phi$,
\begin{equation}
{\tilde S}_{3}^{{\rm dual}}=\int dt dx\left\{-\frac{1}{2{\sqrt{-g}}}\left[\left({\sqrt{-g}}G^0\right)^2-
\left(G^1\right)^2\right]-\frac{1}{2{\sqrt{-g}}}{\lambda}_{++}\left(\sqrt{-g}{G^0}+G^1\right)^2
+\phi{\partial}_{\mu}G^{\mu}\right\}.
\end{equation}
Now treating $\phi$ as a Lagrangian multiplier, we arrive at eq.~(40) once more
and thus deduce the formula of the dual action with the obvious self-duality,
\begin{equation}
S_{3}^{{\rm dual}}=\int dt dx\left\{\frac{1}{2\sqrt{-g}}\left[{\dot \varphi}^2
-\left(\sqrt{-g}{\varphi}^{\prime}\right)^2\right]
-\frac{1}{2{\sqrt{-g}}}{\lambda}_{++}\left({\dot \varphi}-\sqrt{-g}{\varphi}^{\prime}\right)^2\right\}.
\end{equation}
As done in the above two subsections,
the self-duality of $S_{3}$ and $S_{3}^{{\rm dual}}$ may be illustrated by Fig.~3.
\vspace{3mm}
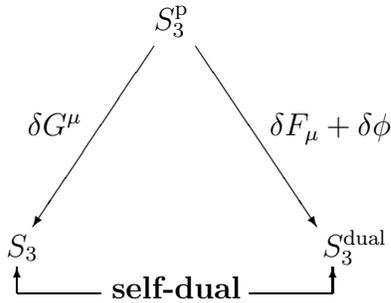
\begin{figure}[htb]
\begin{center}
\setlength{\unitlength}{.7mm}
\begin{picture}(100,50)(0,10)
\put(46,60){$S_{3}^{{\rm p}}$}
\put(46,57){\vector(-2,-3){23}}\put(54,57){\vector(2,-3){23}}
\put(22,40){${\delta G^{\mu}}$}\put(68,40){$\delta F_{\mu}+\delta
\phi$} \put(18,17){${S_{3}}$}\put(78,17){${S_{3}^{{\rm dual}}}$}
\put(20,10){\vector(0,1){5}}\put(80,10){\vector(0,1){5}}
\put(20,10){\line(1,0){16}}\put(64,10){\line(1,0){16}}
\put(38,9){\bf{self-dual}}
\end{picture}
\caption{\small Schematic relation of the actions: the parent action
$S_{3}^{{\rm p}}$ shows that $S_{3}$ and $S_{3}^{{\rm dual}}$ have
self-duality with respect to the chiral boson field.}
\end{center}
\end{figure}

\section{Conclusion and perspective}
In conclusion we emphasize the key point of this paper, that is, the proposal of the noncommutative
extension of the Minkowski spacetime. This newly proposed spacetime
is founded by introducing a proper time from the ${\kappa}$-deformed Minkowski spacetime
related to the standard basis.
It is a commutative spacetime, but contains some information
of noncommutativity encoded from the ${\kappa}$-deformed Minkowski spacetime.
Our performance thus gives the possibility to deal with noncommutativity in a simplified way
within a commutative framework. As a byproduct, a {\em local} field
theory in the ${\kappa}$-deformed Minkowski spacetime can be reduced to
a {\em still local} relativistic field theory in the noncommutative extension of the Minkowski spacetime.
Due to the postulation of the operator
linearization (eq.~(15)) the new spacetime may be regarded as a minimal extension
of the Minkowski spacetime to which the noncommutativity is encoded.
Mathematically, we give a specific map (see eqs.~(4), (5), (7) and (8)):
$({\hat x}^{\mu}, {\hat p}_{\nu}) \rightarrow ({\hat{\cal X}}^{\mu}, {\hat{\cal P}}_{\nu})$,
{\em i.e.} from the ${\kappa}$-Minkowski spacetime to
the extended Minkowski spacetime, and nevertheless the latter spacetime
contains intrinsically the primordial information of the former,
{\em i.e.} the noncommutativity expressed by the finite noncommutative parameter ${\kappa}$.
In applications
one keeps a model Lorentz invariant in the extended Minkowski spacetime, which actually reflects it some
invariance in the ${\kappa}$-Minkowski spacetime related to the standard basis.
This is the method that makes the model comprise noncommutative effects naturally.
Moreover, we notice the connection between the
${\kappa}$-Minkowski spacetime and the flat spacetime with the nontrivial metric eq.~(20), which provides an alternative
way to fulfil noncommutative generalizations for field theory models.
As a primary application of this extended Minkowski spacetime,
three types of formulations of chiral bosons are generalized in terms of this method
to the corresponding noncommutative versions
and the lagrangian theories of noncommutative chiral bosons are acquired and then quantized consistently
by the use of Dirac's method. In addition, the self-duality of the lagrangian theories of noncommutative chiral bosons
is verified in terms of the parent action method. Here we should mention that the self-duality is not
preserved~\cite{s22} when
the chiral boson with the linear chirality constraint is generalized to the canonical noncommutative spacetime.
This implies that the noncommutativity of spacetimes has a complex
relationship with the self-duality of lagrangian theories as has been stated in ref.~\cite{s22}.

In particular,
we note that the proper time $\tau$ (eq.~(16)) contains infinitely many real coefficients, $c_{-n}$ for $n \geq 0$,
which can not be fixed within the framework of the noncommutative
extension of the Minkowski spacetime.
An interesting feature caused by the indefinite coefficients is that the initial value of the proper time,
\begin{equation}
\left.{\tau}\right|_{t=0}=\sum_{n=0}^{+\infty}c_{-n},
\end{equation}
is uncertain even if the convergence requirement of the series of constant terms is added, and
the uncertainty can further extend to any value of the proper time. This may be interpreted to be a kind
of temporal fuzziness that exists in the extended Minkowski spacetime.
Such a fuzziness is compatible with the ${\kappa}$-Minkowski spacetime in which a time and a space
operators commute to the space operator, that is, it originates from the special noncommutativity described in
mathematics by the
${\kappa}$-deformed Poincar{$\acute{\rm e}$} algebra. For noncommutative chiral bosons, nevertheless, we point out that
the fuzziness is different from that discovered~\cite{s22} in the canonical noncommutativity where the similar
phenomenon presents ambiguous left- and right-handed chiralities in the spatial dimension.
In spite of the distinction mentioned, the existence of fuzziness that is closely related to noncommutative spacetimes
might be inevitable
although it is not clear how the noncommutativity brings about fuzziness in the temporal or spatial dimension,
or even probably in multiple temporal and spatial dimensions.

Further considerations focus on the following two aspects. The first
is whether we can give a map from the ${\kappa}$-deformed Minkowski spacetime
related to the bicrossproduct basis~\cite{s7} to a commutative spacetime. This is worth noticing because the
${\kappa}$-Poincar{$\acute{\rm e}$} algebra in the bicrossproduct
basis contains the undeformed (classical) Lorentz subalgebra. Different from the case of the
standard basis, now the difficulty lies in the appearance of entangled terms of
${\hat{p}_0}$ and ${\hat{p}_i}$ in the Casimir operator related to the bicrossproduct basis.
Namely, we try to find, by following the way adopted
for the standard basis or by setting up some new way, such a proper time and such ``proper'' spatial
coordinates as well that the entanglement could be removed. The
second aspect is to enlarge the application of the noncommutative extension of the
Minkowski spacetime by considering any models
that are interesting in the ordinary (commutative) spacetime, in
particular, the models whose noncommutative generalizations
connect with phenomena probably tested in experiments, such as in
ref.~\cite{s26} where the canonical noncommutativity is involved. In
this way we may have opportunities to test the extended
Minkowski spacetime and/or to determine the value of the
noncommutative parameter by comparing theoretical results with
available experimental data. A quick and direct consideration is to investigate chiral $p$-forms~\cite{s18,s19}
and the interacting theory of chiral bosons and gauge
fields~\cite{s27} in the noncommutative extension of the Minkowski spacetime, and results will be given separately.


\vspace{10mm}
\noindent
{\bf Acknowledgments}
\vspace{5mm}

\noindent
The author would like to thank H.J.W. M\"uller-Kirsten of the University of Kaiserslautern for helpful discussions and
H.P. Nilles of the University of Bonn for warm hospitality.
This work was supported in part by the DFG (Deutsche Forschungsgemeinschaft), by the National Natural
Science Foundation of China under grant No.10675061,
and by the Ministry of Education of China under grant No.20060055006.

\end{document}